\documentclass[trackchanges,twocolumn]{aastex701}

\usepackage{amsmath}

\usepackage{xcolor}
\usepackage{soul}

\begin{document}

\title{The Edge of the Warm Corona in Active Galactic Nuclei}

\author[orcid=0009-0004-6142-9668,sname='Biswas']{Parikshit Partha Biswas}
\affiliation{Nicolaus Copernicus Astronomical Center Polish Academy of Sciences, Bartycka 18, 00-716 Warszawa, Poland}
\email[show]{pbiswas@camk.edu.pl}  

\author[orcid=0000-0002-5275-4096,sname='Różańska']{Agata Różańska}
\affiliation{Nicolaus Copernicus Astronomical Center Polish Academy of Sciences, Bartycka 18, 00-716 Warszawa, Poland}
\email{agata@camk.edu.pl}  

\author[orcid=0000-0003-0826-9787,sname='Lančová']{Debora Lančová}
\affiliation{Nicolaus Copernicus Astronomical Center Polish Academy of Sciences, Bartycka 18, 00-716 Warszawa, Poland}
\affiliation{Research Center for Computational Physics and Data Processing, Institute of Physics, Silesian University in Opava, Bezru\v{c}ovo n\'{a}m. 13, 746 01 Opava, Czech Republic}
\email{debora.lancova@physics.cz}  

\author[orcid=0000-0001-5325-7705,sname='Gronkiewicz']{Dominik Gronkiewicz}
\affiliation{Space Research Center Polish Academy of Sciences, Solar Physics Division, Kopernika 11, 51-622 Wrocław, Poland}
\email{}

\author[orcid=0000-0003-4971-2538,sname='Wojaczyński']{Rafał Wojaczyński}
\affiliation{Nicolaus Copernicus Astronomical Center Polish Academy of Sciences, Bartycka 18, 00-716 Warszawa, Poland}
\email{rwojaczynski@camk.edu.pl}

\author[orcid=0000-0002-0333-2452,sname='Zdziarski']{Andrzej Zdziarski}
\affiliation{Nicolaus Copernicus Astronomical Center Polish Academy of Sciences, Bartycka 18, 00-716 Warszawa, Poland}
\email{aaz@camk.edu.pl}

\author[orcid=0000-0002-4622-4240,sname='Bianchi']{Stefano Bianchi}
\affiliation{Università degli Studi Roma Tre - Via della Vasca Navale 84, I-00146 – Rome, Italy}
\email{stefano.bianchi@uniroma3.it}  

\begin{abstract}
 {   A model of warm, dissipative corona on top of an accretion disk is investigated, to explain the soft X-ray excess (SXE) emission observed in active galactic nuclei (AGN). The final ray-traced spectrum, computed through photoionization code with Comptonization, accounts for both: emitted and reflected components. As a result, the strength of the SXE depends on the radial extend of the warm corona, i.e., its outer edge.} We show that the {   radial} transition from warm corona to disk {   atmosphere} is governed by line cooling. By moving away from a central black hole Compton cooling is {   gradually} replaced by line cooling, and the temperature drops sharply {   in vertical direction} at a narrow iron ionization front, where iron ions Fe\,\textsc{xxv}--Fe\,\textsc{xxvi} recombine to  Fe\,\textsc{xvii}. {   Notably, this transition takes place at a particular value of internal heating $\approx 1.6 \times 10^{-23}$\,erg\,cm$^{3}$\,s$^{-1}$}, regardless of the Eddington ratio. We show, that {   at the given metal abundance}, the SXE emitting area is larger for a higher accretion rate, in agreement with recent findings in changing look AGN. The overall emission spectrum from a warm corona is featureless, and the only iron K$_{\alpha}$ line comes from a reflection. The line is stronger and more gravitationally distorted for a lower accretion rate, which is a direct result of the size of the warm corona. Furthermore, by increasing {   the iron abundance}, the {   iron} ionization front moves closer towards black hole, and a stronger iron line complex is emitted locally.  
\end{abstract}


\section{Introduction}

Active galactic nuclei (AGN)  {with} a supermassive black hole (SMBH) in their centers are strong X-ray emitters.  {This emission may originate directly from a hot compact corona within a few gravitational radii of the central SMBH} \citep[][and references therein]{2017fabian}. Such a hot corona, if thermalized, reaches temperatures of the order of 100 keV, and its spectrum is well modeled by a power-law, produced by Compton up-scattering of soft photons. Nevertheless, more than half of AGN exhibit a deviation from this power-law shape below $\approx 2$~keV, commonly known as soft X-ray excess (SXE). The SXE has been interpreted either as an emission from a warm, optically thick,  Comptonized structure with a temperature of $\sim 10^7$ K, i.e. warm corona \citep[e.g.,][and references therein]{1998Magdziarz,2004Gierlinski,2012Done,2013Petrucci}, or as a blurred ionized reflection from the accretion disk close to the SMBH \citep[e.g.,][and references therein]{1993Ross,2005Ross,2014Garcia}.

{   Based on} X-ray data analysis, both models fit equally well \citep{2019Garcia}. {   For some sources, a warm corona above an accretion disk is favored over blurred ionized reflection \citep{2016Boissay}, while in other sources, blurred ionized reflection is preferred} \citep[e.g.,][]{2018Mallick,2021Xu}. {   Although there is no need to use both models to explain just the SXE curvature seen in the data, the model \texttt{recXcor} combines the reflection with warm Comptonization \citep{2020Ballantyne,2022Xiang}, and it was used to fit the broadband X-ray spectrum in order to explain SXE together with the iron line profile. 
The \texttt{recXcor} model has been tested on several sources \citep{2022Xiang,2024Porquet,2025Porquet}, and shown that both Comptonization and reflection have been required, but they were most significant in the different energy ranges, i.e Comptonization in SXE and reflection in the iron line region.} 

{   On the other hand, it has been theoretically predicted that the existence of} an optically thick, Compton-cooled warm corona {   on the top of an accretion disk in radiative equilibrium} requires an additional dissipative process that heats all parts of the warm zone \citep[][]{2015Rozanska}. The above geometry of an accretion flow  {was reproduced} by recent MHD simulations, which show the existence of a warm corona structure for the case of an AGN accretion on the black hole of $10^8\ {\rm M_{\odot}}$ \citep[see,][and references therein]{2026Nagele}.  {Hence, a two-slab model, in which a warm and dissipative corona lies on top of an accretion disk, has emerged as the most promising scenario {   for reproducing the SXE} \citep{2018Petrucci,2020Ballantyne,2023Gronkiewicz,2024Ballantyne,2024Palit}}.

{   Thus, the dissipative warm corona model as an outcome of the radiative transfer code \texttt{TITAN/NOAR} \citep{2020Petrucci}}, has been fitted to the 82 data sets of 21 AGN \citep{2024Palit}. As a result of this analysis, it has been reported 
 that the strength of the SXE (ratio of the soft X-ray flux to the extrapolated hard X-ray continuum) and the warm corona radius correlate with the Eddington ratio, assuming the same cylindrical geometry as an accretion disk. The same authors found that the warm region spans a broad range, from 7 to 408 gravitational radii, and that the warm corona is more extended at higher accretion rates. Such a positive correlation indicates that a warm corona size directly depends on the Eddington ratio \citep{2024Ballantyne,2024Palit}, and may in turn initiate changes in the accretion flow, significantly affecting its geometry. {   This latter conclusion is additionally} motivated by recent discoveries showing that the warm corona appears during the spectral state transition observed in so-called changing-look AGN (CLAGN), when the source undergoes re-brightening \citep{2025palit,2026Jana}. 

In this work, we {   numerically put constraints on} the geometrical extent of the warm corona considering the dominant heating and cooling processes occurring in the gas heated to temperatures of the order of millions of Kelvin. {   To solve the vertical structure of a warm, Comptonized disk atmosphere} we use the photoionization code \texttt{TITAN/NOAR} \citep{2003Dumont}. These solutions are combined with the radial structure of an accretion disk defined by \citet{2022Xiang} followed by \citet{2026Biswas}.  {As the distance from the SMBH increases, we find a sudden drop in the temperature, consistent with a switch in the dominant cooling mechanism operating in the dissipative and illuminated disk atmosphere. We demonstrate that such a drop corresponds directly to the replacement of Compton cooling by line cooling, and we identify the radius of this transition with the outer edge of the warm corona.}
With this definition of the transition {   from the inner disk atmosphere composed of optically thick, Comptonized slab to the disk atmosphere where Compton cooling is negligible}, we demonstrate that {   the size of} the warm-corona-emitting region increases with increasing Eddington ratio, in agreement with recent observations \citep{2025palit,2026Jana}. 
Next, following the procedure outlined in \citet{2026Biswas}, we put observational predictions on the emission spectra from the dissipative warm corona.
For this purpose, we use 
publicly available ray {-}tracing code \texttt{GYOTO}\footnote{\href{https://gyoto.obspm.fr/index.html}{https://gyoto.obspm.fr/index.html}} \citep{2011Vincent},
for general-relativistic ray-tracing of the source emission as seen by a distant observer. 

This paper is organized as follows: {   Sec.~\ref{subsec:def_wc} describes physically justified geometry of a warm corona}; Sec.~\ref{sec:method} presents the computational methodology, the numerical codes \texttt{TITAN+NOAR+GYOTO} (hereafter \texttt{TiNoGyo}), and our parameter setup. Sec.~\ref{sec:res} shows the results displaying the dependence of the size of the warm corona on the Eddington ratio and the iron abundance. Sec.~\ref{sec:discon}  contains discussion and conclusions.

\section{Outer Edge of the Warm Corona}
\label{subsec:def_wc}

\begin{figure}
    \centering
    \includegraphics[width=0.99\linewidth]{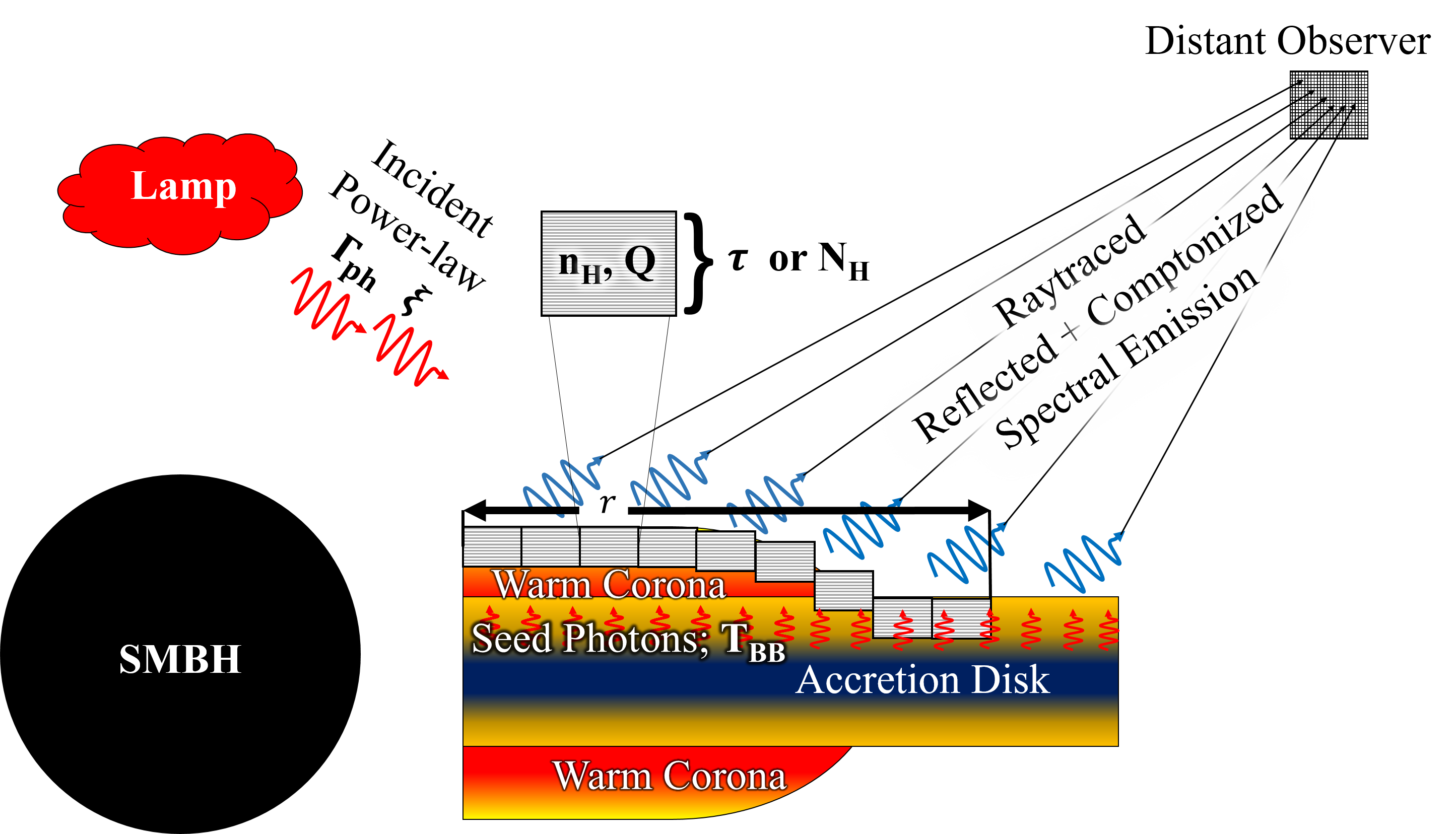}
    \caption{   Schematic of the  \texttt{TiNoGyo} calculation setup.
    The gray box marks the slab of optical depth $\tau$ used for radiative transfer calculations.
    The lamp provides incident power-law with photon index $\Gamma_{\rm ph}$ that heats up the atmosphere. 
   The illuminated flux defines the ionization parameter $\xi(r)$, which, along with other parameters $n_{   H}(r)$, $Q(r)$, and $T_{\rm BB}(r)$, depends on the radius according to the formulas given by Eqs.~\ref{eq:D}--\ref{eq:inputs}. 
    A distant observer sees both the reflected and Comptonized spectrum, integrated over the emitting surface.} 
    \label{fig:schematic_dig}
\end{figure}

{   We consider a warm corona to be an optically thick, dissipative atmosphere located on top of a standard accretion disk \citep{1973Shakura}, as depicted schematically in Fig.~\ref{fig:schematic_dig} by gray boxes. The assumption of a standard disk is needed to account for the total energy deposition caused by accretion at a given radius $r$: 
\begin{equation}
\begin{split}
D(r) = {} & 6.89\times10^{27}\left(\frac{\eta}{0.1}\right)^{-1}\left(\frac{M}{{\rm M}_\odot}\right)^{-1}\\
& \times\lambda_{\rm Edd}\,r^{-3}J(r)\ {\rm erg\,cm^{-2}\,s^{-1}},
\end{split}
\label{eq:D}
\end{equation}
where $M$ is the SMBH mass, $\lambda_{\rm Edd}$ is an accretion ratio (i.e., accretion rate in units of the Eddington accretion rate), and  $\eta$ is the efficiency of accretion.} 
The stress-free inner boundary condition is given by $J(r)=1-(r_{\rm ISCO}/r)^{1/2}$ where $r_{\rm ISCO}$ is the innermost stable circular orbit.
At each radius, we assume the same division of this energy budget among three components: 
the hot corona in the lamp post geometry (marked in red in our diagram) dissipates a fraction $f_X=0.2$ of the total energy, the warm corona dissipates a fraction $f_W=0.4$, and the rest $1-f_X-f_W=0.4$ is generated in the disk itself.
This arbitrary division is needed in order to solve the radiative interaction and cooling channels providing to the overall source emission in the soft X-ray band. 

As a result of many observational studies, we expect the warm corona to emit a Comptonized spectrum producing SXE \citep{1998Magdziarz,2013Petrucci}. Therefore, we consider a warm corona to be a part of the accretion disk atmosphere that is illuminated from outside by hard X-rays from the lamp, and from below by disk seed photons with a blackbody temperature. Thus, in the vertical direction, a warm corona and an accretion disk form a two-slab geometry \citep{1994Svensson}, in which the gas number density of the disk atmosphere is taken to be a factor of $10^{3}$ below the mid-plane density of a radiation-pressure-dominated disk:
\begin{equation}
\begin{split}
n_{\rm H}(r) = {} & 2.4\times10^{15}\left(\frac{\eta}{0.1}\right)^{2}\left(\frac{\alpha}{0.1}\right)^{-1}\left(\frac{M}{{\rm M}_\odot}\right)^{-1}\\
& \times\lambda_{\rm Edd}^{-2}\,r^{3/2}J(r)^{-2}\ {\rm cm^{-3}},
\end{split}
\label{eq:nH}
\end{equation}
where $\alpha=0.1$ is the viscosity parameter, and $r$ is in units of $r_{\rm g}=GM/c^2$, where $G$ and $c$ are the gravitational constant and the speed of light, respectively.

{   The temperature of the heated gas, and therefore an outgoing emission, can be established by solving the radiative transfer through the atmosphere illuminated by hard X-ray flux from the lamp above the SMBH. The lamp luminosity is calculated as $L_X=f_X\int_{r_{\rm ISCO}}^{r_{\rm c}}D(r)\,dS(r)$, where $r_{\rm c}=10\,r_{\rm g}$ and $dS$ is the proper area of a disk element. Following \citet{2022Xiang}, we adopt the illumination pattern $\epsilon(r)$ of a lamp at height $h=20\,r_{\rm g}$, including light bending and the energy shift, normalized to $\int\epsilon\,dS=1$. Thus, the radial distribution of the flux from the lamp corresponds to $F_X(r)=L_X\,\epsilon(r)$ \citep[see][for full expression]{2026Biswas}.}
We compute radiative transfer at each radius with the use of \texttt{TITAN/NOAR} photoionization code, which accounts for all physical heating and cooling processes, such as internal heating, Comptonization, Bremsstrahlung, photoionization, recombination, and 4141 major line transitions {   on major heavy elements} \citep[][and references therein]{2026Biswas}. 
 {At each depth of optically thick atmosphere, the temperature is set by the nLTE energy balance between all cooling and heating processes. As an outcome, we obtain the local structure of the warm corona with all values of energy and radiation outputs.}

%\annot{I'm leaving this to explain Parik that this is not true: 
%What matters is therefore not only the amount of heating, but which process balances it. If $Q$ is too large, the gas overheats and becomes fully ionized, and it then radiates a bremsstrahlung continuum reaching well past the soft X-ray band instead of the smooth excess that is observed} \citep{2020Ballantyne_b}.

{   It was already demonstrated by \citet[][see Fig.2]{2026Biswas} that since dissipated heating decreases as the radius increases (Eq.~\ref{eq:D}), and gas density increases with the radius (Eq.~\ref{eq:nH}), the temperature of a warm corona also decreases with distance from the SMBH.} 
This effect is reinforced by values of other parameters, such as ionization degree $\xi(r)$, dissipative heating in the atmosphere $Q(r)$, and temperature of seed photons $T_{\rm BB}(r)$, needed to keep the corona warm. All of them are decreasing with radius, according to formulae:
\begin{equation}
\begin{aligned}
\xi(r) &= \frac{4\pi F_{X}(r)}{n_{\rm H}(r)},\\
Q(r) &= \frac{f_W\,\sigma_{\rm T}\,D(r)}{\tau\,n_{\rm H}(r)},\\
T_{\rm BB}(r) &= \left[\frac{(1-f_X-f_W)\,D(r)}{\sigma}\right]^{1/4},
\end{aligned}
\label{eq:inputs}
\end{equation}
where {   $\sigma_{\rm T}$ is Thomson cross section, $\tau$ depicts total optical depth of atmosphere,} and $\sigma$ is the Stefan--Boltzmann constant.

 {Since the warm corona requires both a sufficiently heated disk atmosphere and Compton cooling as the dominant channel, it can only survive over a limited range of radii.
If $Q(r)$ is too large, the gas becomes fully ionized, giving rise to a medium that can be cooled efficiently by Compton scattering.
If instead the cooling becomes too efficient for a given $Q(r)$, the disk atmosphere settles at a lower temperature, where other cooling channels become important. 
Because the model is calculated self-consistently, we can determine the radius at which Comptonization ceases to be important, and we identify this radius with the outer edge of the warm corona $r_{   edge}$. We show that beyond this edge, the disk atmosphere is dominated by line cooling rather than by Compton cooling, and does not account for the SXE strength defined as:}
\begin{equation}
    S_{\rm SXE}(r) = \frac{\int^{2}_{0.5} F^{\rm CE}(r) (dE/E)}{\int^{2}_{0.5} F^{\rm tot}(r) (dE/E)},
    \label{eq:SXE_strength}
\end{equation}
where $F^{\rm CE}(r)$ is the monochromatic flux of the Comptonized emission and $F^{\rm tot} (r)$ is the monochromatic total flux, i.e., Comptonized plus reflected spectrum, both being an outcome of 
\texttt{TITAN/NOAR} at each radius. 
{   In the sections below, we present the size of the warm corona for different accretion rates and iron abundance. We show how it affects the final outgoing emission and SXE strength, and we clearly indicate the reason why the warm corona ends. 
}

%%%%%%%%%%%%%%%%%%%%%%%%%%%%%%%%%%%%%%%%%%%%%%%%%%%%%%%%%%%%%%
\section{Methodology}
\label{sec:method}

We study both the radial and vertical structure of an accretion flow, but both dimensions are calculated separately and are linked by their input parameters {   given by Eqs.~\ref{eq:D}-\ref{eq:inputs}}. The vertical stratification is obtained with the photoionization code \texttt{TITAN}, while the radial stratification is defined by analytical formulae adapted from \citet{2022Xiang} and discussed in detail by \citet{2026Biswas}. We further couple \texttt{TITAN} with \texttt{NOAR} to include Comptonization shifts in the spectra. Finally, the relativistic effects are added by feeding the local \texttt{TITAN+NOAR} spectra to the ray-tracing code \texttt{GYOTO}. Together, they form what we call here the \texttt{TiNoGyo} model.

\subsection{\texttt{TiNoGyo}}\label{subsec:titan}

\texttt{TITAN} is a one-dimensional photoionization code that solves the radiative transfer equation through a plane-parallel atmosphere, allowing illumination from both sides. As a result, we obtain the ionized vertical structure of the  {disk atmosphere} and the emerging angle-dependent spectrum in continuum and lines. The local stratification of the computed  {disk atmosphere} is depicted in Fig.~\ref{fig:schematic_dig} {   by the gray box}.
We use the newest version of \texttt{TITAN} that allows us to take into account additional internal heating of the atmosphere \citep{2003Dumont,2020Petrucci}, which is needed for a self-consistent treatment of a dissipative warm corona.

For the hot corona, we adopt a lamp-post geometry where the  {outer disk atmosphere} is illuminated by a hard X-ray power-law spectral distribution with photon index $\Gamma_{\rm ph}$, as shown in Fig.~\ref{fig:schematic_dig}. From the bottom,  {the atmosphere} is supplied by seed photons with a blackbody temperature $T_{\rm BB}(r)$, computed from the energy flux dissipated in the disk. At a given radius, we assume the gas number density $n_{\rm H}(r)$ and the additional internal heating $Q(r)$ to be constant in the vertical direction throughout the  {atmosphere}, as discussed in \citet{2015Rozanska,2026Biswas}. We also assume the total hydrogen column density $N_{\rm H}$, which translates to optical depth $\tau$, to be constant at each radius throughout the  { disk atmosphere}.

In \texttt{TITAN}, the thermal and ionization state in each zone is computed by assuming a steady state, such that the local balance between ionization and recombination of ions, excitations and de-excitations, local energy balance (heating due to absorption and cooling due to local emission), and total energy balance (equality between inward and outward fluxes) are maintained \citep{2000Dumont}. \texttt{TITAN} takes into account all relevant physical processes to cover the large range of temperature which may be reached in each zone. The ionization equilibrium equations include radiative ionization by continuum and line photons, collisional ionization and recombination, radiative and dielectronic recombination, charge transfer by H and He atoms, the Auger effect, and ionization by {   hard} X-ray photons. \texttt{TITAN} includes 10 elements (H, He, C, N, O, Ne, Mg, Si, S, and Fe) as its gas composition. Hence, the abundances of these elements can be treated as an input parameter, with the default values set to Solar abundance \citep{1989Andres}. 

The output from \texttt{TITAN} provides the gas number density, temperature, and ionization structure of the matter. These are used as input for \texttt{NOAR}, which produces an angle-dependent Comptonized emission of both: {   thermal emission of a warm atmosphere and the reflected component of the spectrum}. 
{   We would like to point out a minor discrepancy in our models}.
The Monte Carlo Comptonization code \texttt{NOAR} differs from \texttt{TITAN} in how it sets a lower boundary condition on the flux injection from the underlying disk. While \texttt{TITAN} can include appropriate differential boundary conditions on the true intensity from below, \texttt{NOAR} allows only for {   constant} photon flux injection. 
This is because the photoionization code solves the classical radiative transfer equation, while \texttt{NOAR} uses Monte Carlo photon tracking. Therefore, the final emission models, presented in this paper, 
{   may not fully reproduce the proper spectral slope}, but they are {   correct} to illustrate the relative differences in SXE strength.

Finally, to build a complete model, we included the relativistic effects in the spectrum seen by a distant observer. To do so, we use the Comptonized spectral output from \texttt{NOAR} and feed it into the general-relativistic ray-tracing code \texttt{GYOTO}. We use the Directional Disk model in \texttt{GYOTO}, which requires a radially stratified angle-dependent spectrum as its input. \texttt{GYOTO} uses a ray-tracing technique to trace photon paths and includes all relativistic effects {   caused by the SMBH }\citep[see][for further details]{2011Vincent,2016Vincent,2026Biswas}.

\subsection{Setup} \label{subsec:setup}

{   Given the complexity of the model, to study the physical properties of the warm corona systematically, we fix some of the parameters at constant values.} In particular, the black hole is not rotating, and its mass is set to $M=10^8\, {\rm M_{\odot}}$. The power-law spectral index is kept constant at $\Gamma_{\rm ph}=2$, and the column density is chosen as $N_{\rm H} = 6.1639 \times 10^{24}\, {\rm cm^{-2}}$ which corresponds to the  {disk atmosphere's} {   effective} optical depth $\tau \approx 5$. 
{   The total optical depth is needed as a lower boundary condition in photoionization calculations. It does not correspond to the optical depth of the warm corona inferred from observations. This is because \texttt{TITAN} calculations are made locally, while the total spectrum seen by a distant observer is integrated over the emission surface. By increasing the value of $\tau$, 
dissipated heating $Q(r)$ is getting lower according to eq.~\ref{eq:inputs}. Our choice is a compromise between efficient heating of the atmosphere and being optically thick to reproduce Comptonized warm corona.}

{   The radial grid extends from $r_{\rm ISCO}+1\,r_{\rm g}$ }
to just beyond the outer edge of the warm corona (Sec.~\ref{sec:res}). All radii in our model are expressed in units of $r_{\rm g}$. In this paper, we focus on the location of the outer edge of the warm corona, governed by physical processes in the ionized gas. To emphasize {   how the position of the ionization front changes, we consider three different Eddington ratios, given in Tab.~\ref{tab:wc_edd}, with a constant accretion efficiency of $\eta = 0.057$ for a non-spinning black hole, needed in Eq.~\ref{eq:D}.}

\begin{figure}[]
    \centering    \includegraphics[width=0.99\linewidth]{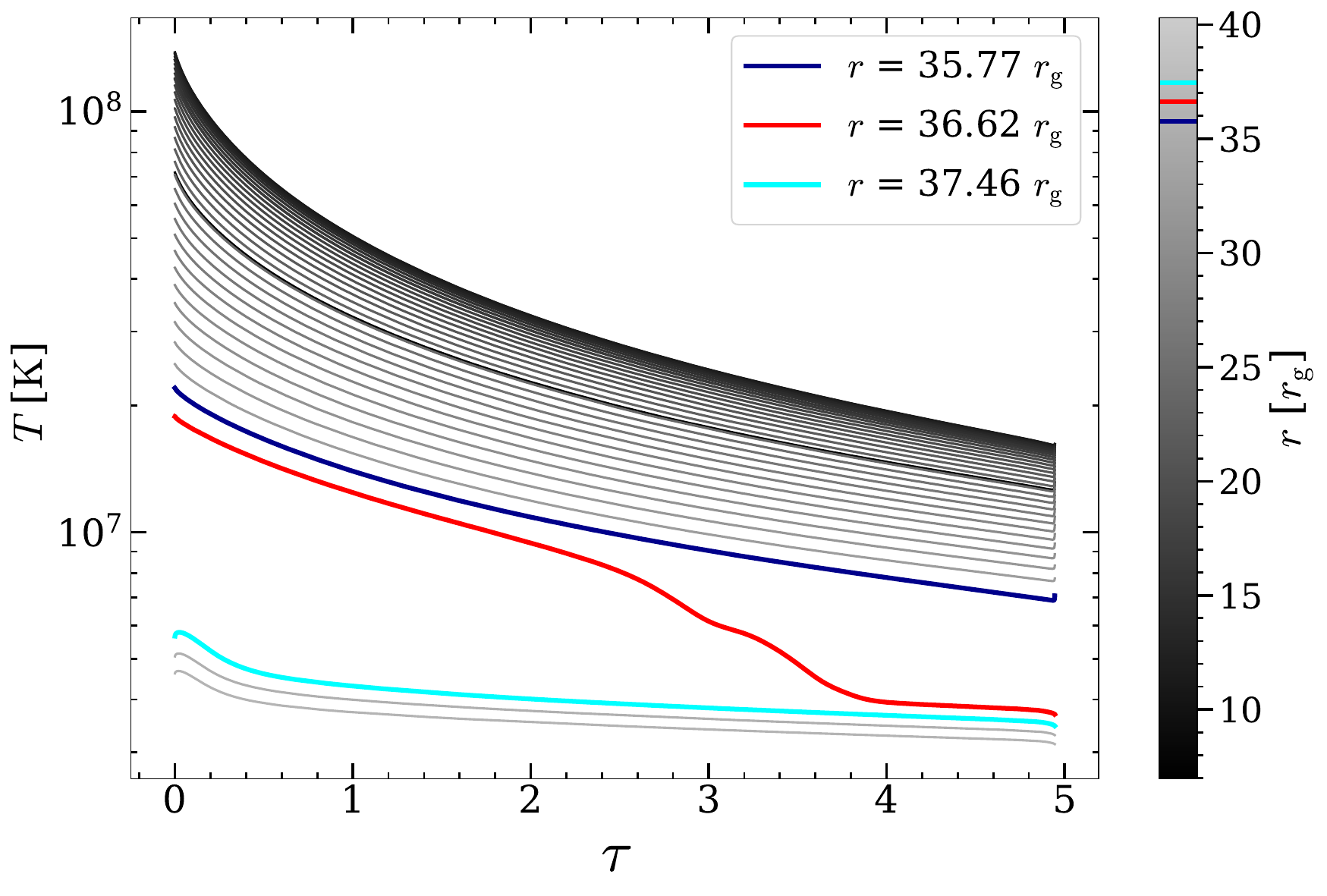}
    \caption{Temperature as a function of optical depth shown at different distances from the SMBH, for the case of $\lambda_{\rm Edd} = 0.1$ and Solar abundance. The grayscale gradient shows the radial structure with the scale given in the side bar.
    The transitional region is depicted and marked in color.}
    \label{fig:temp_spin_0}
\end{figure}

\begin{figure*}[]
    \centering
\includegraphics[width=0.99\linewidth]{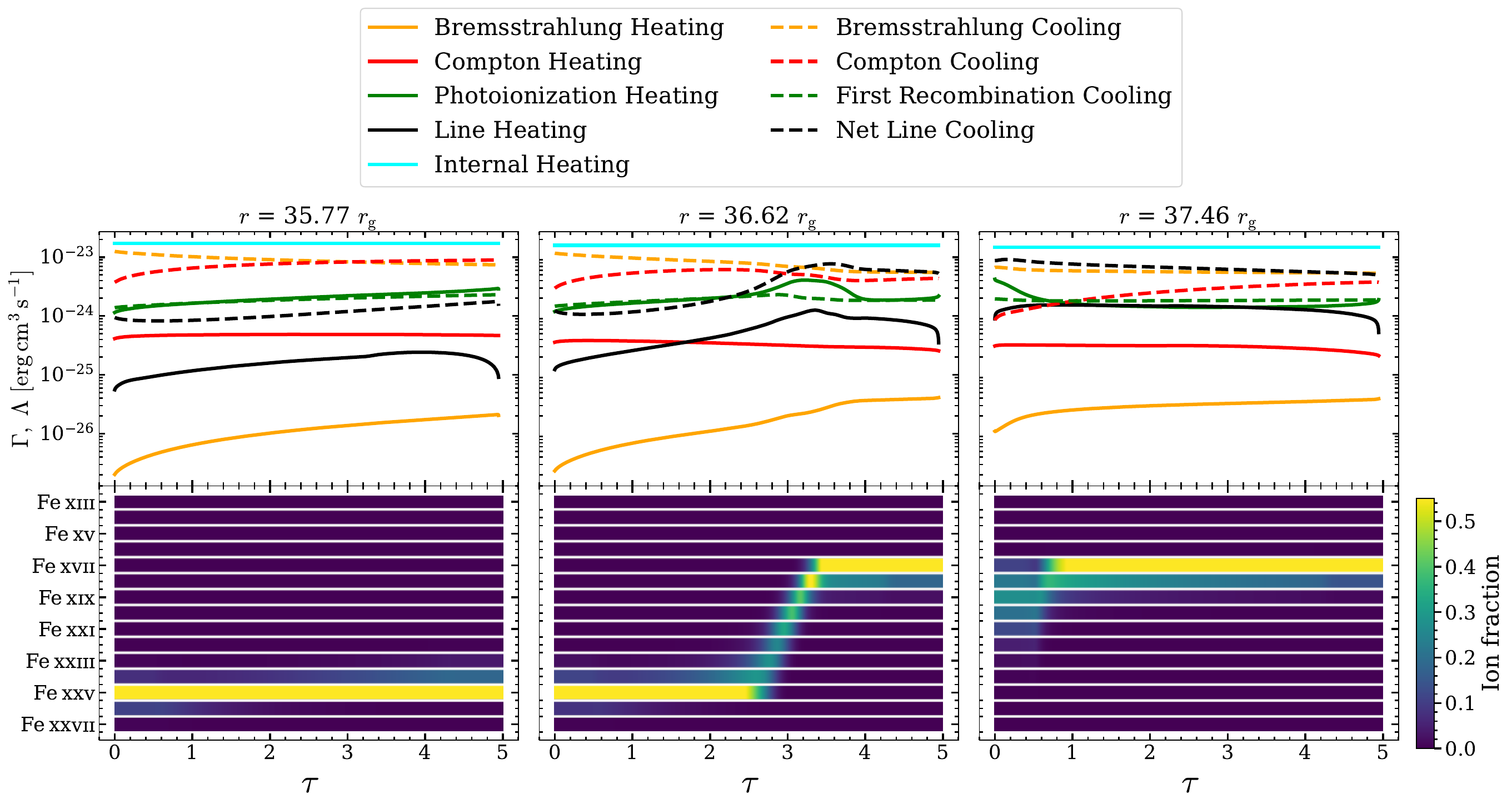}
    \caption{The top panels depict the heating ($\Gamma$) and cooling ($\Lambda$) rates versus the optical depth {   for the same case as Fig.~\ref{fig:temp_spin_0}}. Different physical processes, given in the box above the panels, are presented by the solid and dashed lines, respectively. 
    The bottom panels show the fraction of iron ions at different ionization states versus optical depth.
    {   Each column of panels corresponds to one radius, indicated above the plots. Strong ionization front of iron is visible within the transitional region.}
    }
    \label{fig:hc_spin_0}
\end{figure*}

{   In addition}, we examine how the iron content affects the warm corona structure. We adopt the standard astronomical logarithmic abundance scale defined as:
\begin{equation}
    \log(\varepsilon_{\rm Fe}) =  \log\left( \frac{n_{\rm Fe}}{n_{\rm H}} \right) + 12,
\end{equation}
where $\log(\varepsilon_{\rm Fe})$ is the logarithmic absolute abundance of iron, $n_{\rm Fe}$ is the iron number density, and the additive constant 12 represents the normalization of the abundance scale, in which hydrogen is assigned a reference abundance of $\log(\varepsilon_{\rm H})=12$ \citep{1989Andres}. We consider three cases from Solar to Super Solar abundance, as depicted in Tab.~\ref{tab:Fe_abund}. While varying the Eddington ratio, we kept the iron content as Solar, and for the variation of the iron abundance, we fixed the accretion rate as $\lambda_{\rm Edd} = 0.1$.

 {
The Eddington ratio enters Eqs.~\ref{eq:D}--\ref{eq:inputs} only as a multiplicative factor. Changing $\lambda_{\rm Edd}$ therefore rescales each input profile by a constant at all radii without changing its radial shape. Hence, at fixed $r$,
\begin{equation}
\begin{gathered}
n_{\rm H}(r)\propto\lambda_{\rm Edd}^{-2},\quad D(r),\,F_X(r)\propto\lambda_{\rm Edd},\\
T_{\rm BB}(r)\propto\lambda_{\rm Edd}^{1/4},\quad \xi(r),\,Q(r)\propto\lambda_{\rm Edd}^{3}.
\end{gathered}
\label{eq:scaling}
\end{equation}
The steep scaling of $\xi(r)$ and $Q(r)$ arises because the dissipated flux grows linearly with $\lambda_{\rm Edd}$, while the inner disk becomes less dense, $n_{\rm H}\propto\lambda_{\rm Edd}^{-2}$. The internal heating is thus proportional to the dissipated energy flux per gas number density, $Q(r)\propto D(r)/n(r)_{\rm H}\propto\lambda_{\rm Edd}^{3}\,r^{-9/2}J(r)^{3}$, and $\xi(r)$ is the analogous quantity for the incident X-ray flux.
}

%%%%%%%%%%%%%%%%%%%%%%%%%%%%%%%%%%%%%%%%%%%%%%%%%%%%%%%%%%%%%%
\section{Results}
\label{sec:res}

{   The main result of our paper is the identification of the transition zone where the Compton-cooled disk atmosphere, attributed to the warm corona, ends and the line-cooled disk atmosphere begins, marked by a drop in gas temperature.}
We first examine in detail the case of Solar abundance and $\lambda_{\rm Edd} = 0.1$. 

 { The temperature structures with optical depth at several distances from the SMBH are presented in Fig.~\ref{fig:temp_spin_0}. In most cases, they are flat, except at a radius of $r=36.62\, r_{\rm g}$, the temperature drops by a factor of $\approx 2.5$, from $T\approx 10^7\, {\rm K}$ to $T\approx 4\times10^6\, {\rm K}$, within $2\leq \tau \leq 4$}. To understand this drop, we examine the gas state in detail at three radii marked in dark blue, red, and cyan in Fig.~\ref{fig:temp_spin_0}. For these respective radii, in Fig.~\ref{fig:hc_spin_0}, we plot all  heating ($\Gamma$) and cooling ($\Lambda$) rates {   operating in the atmosphere} 
and the corresponding iron ion's fraction, i.e., the population of ions at a given ionization state relative to the total iron content, both as a function of optical depth. Together, Fig.~\ref{fig:temp_spin_0} and Fig.~\ref{fig:hc_spin_0} show that the temperature drop aligns with the dominant ion shifting from Fe\,\textsc{xxv} to Fe\,\textsc{xvii}, alongside a rise in line cooling rate.
Fe\,\textsc{xvii} (Ne-like, with closed ${\rm 2p^6}$ shell) and Fe\,\textsc{xxv} (He-like, with closed ${\rm 1s^2}$ shell) are unusually stable configurations, and removing an electron from either costs substantially more energy than removing one from any of the open-shell ions between them, Fe\,\textsc{xviii} to Fe\,\textsc{xxiv} \citep{2019Deprince}.

\begin{figure*}
    \centering    \includegraphics[width=0.99\linewidth]{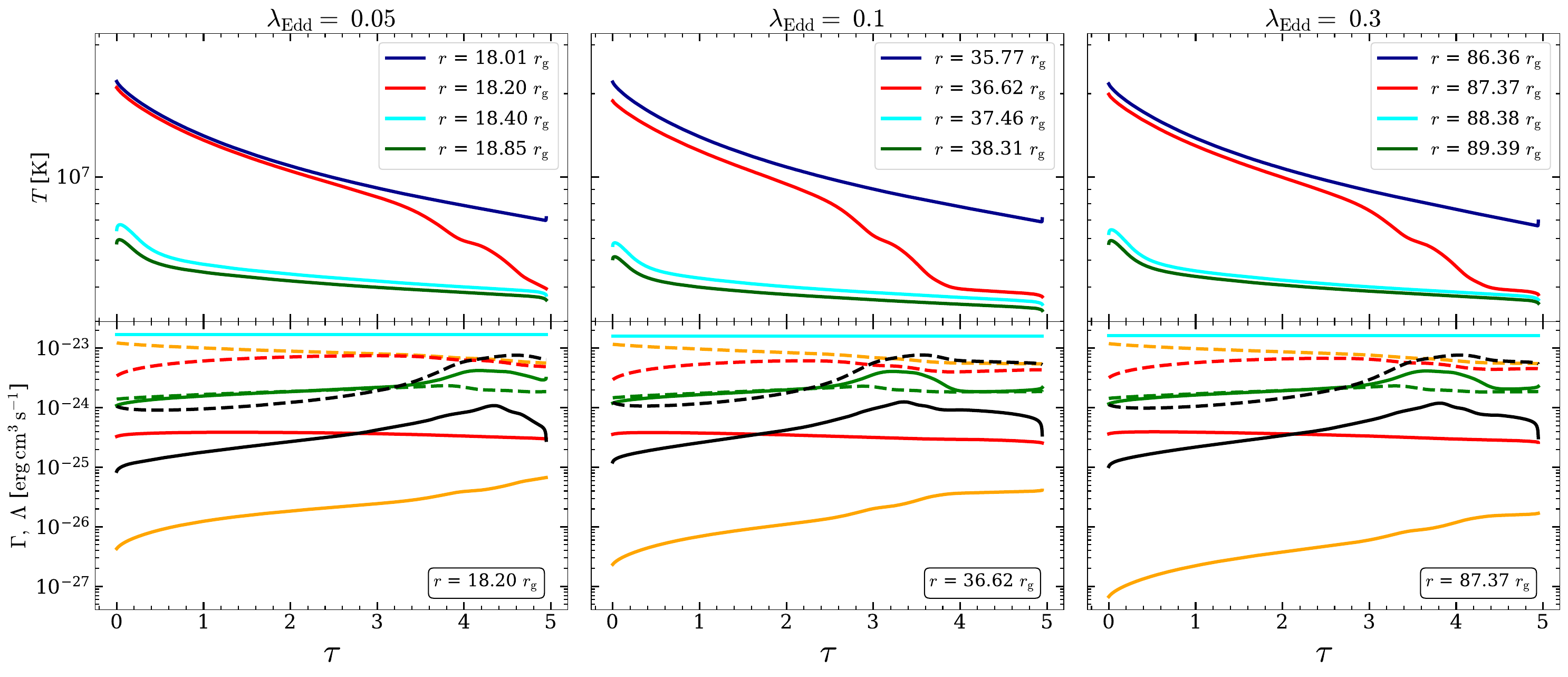}
    \caption{The top panels show the variation of temperature with the optical depth for different $\lambda_{\rm Edd}$ and radial distances, as indicated by labels. For the radii where temperature drop occurs (red line), the corresponding variations of the heating and cooling rates are {   plotted} in the bottom panels. {   The meaning of $\Gamma$ and $\Lambda$ is the same as in Fig.~\ref{fig:hc_spin_0}.}}
    \label{fig:edding_ratio}
\end{figure*}

As the iron ions move towards a lower ionization state from Fe\,\textsc{xxv} to Fe\,\textsc{xvii}, the L-shell starts filling with electrons. The hot free electrons in the warm corona transfer part of their kinetic energy to L-shell electrons through collisions, decreasing the kinetic energy of the free electrons. These collisions excite the L-shell electron to a higher level in the same iron atom. Since the excited electron remains bound to the same ion, it then de-excites and releases a line photon, cooling the gas \citep{2006Osterbrock}. Such a line cooling mechanism is the main driver responsible for radiative cooling of the warm corona and transitioning it to a  {line-cooled disk atmosphere}. We note that for ions like Fe\,\textsc{xxv} (${\rm 1s^2}$) and Fe\,\textsc{xxvi} (${\rm 1s^1}$), line cooling still exists through excitation of the K-shell electrons, but these electrons are far more tightly bound, so this channel is not dominant. Hence, in Fig.~\ref{fig:hc_spin_0}, {   we see that as the iron ions recombine from a higher to a lower ionization state at $\tau \approx 2.5$,} where line cooling becomes more efficient and starts to dominate, resulting in a sudden decrease in temperature.

 {
As discussed in Sec.~2, if Compton cooling is the dominant cooling process, the disk atmosphere produces SXE and constitutes a warm corona. Beyond a certain radius, Compton cooling is no longer dominant, and the SXE emission weakens. This provides a natural way to define the transition region, over which the disk atmosphere changes from a warm corona to a line-cooled zone with the assumption of $\tau = 5$. In the case of a 0.1 accretion rate and Solar iron abundance, this happens at $36.62\,r_{\rm g}$, where we first notice the temperature drop, and we define this radius as the edge of the warm corona. 
 A similar radial restriction was found by 
 \citep{2024Kawanaka}, who showed that the warm corona temperature scales with the underlying disk temperature and therefore falls steeply with radius, so that a warm corona can be sustained only close to the black hole.}

\subsection{Effect of the Eddington Ratio on the Warm Corona Edge} \label{sec:Edding_ratio}

Next, we repeat the same analysis for two other values of the Eddington ratio,  $\lambda_{\rm Edd} = 0.05$ and 0.3. In the top panels of Fig.~\ref{fig:edding_ratio}, we show the temperature structure as a function of the optical depth at different radii, from which we identify the transitional region.  {The temperature drop occurs in each case of Solar iron abundance, and the location of the warm corona moves outward with increasing $\lambda_{\rm Edd}$, from $r_{\rm edge}$ of $18.2\,r_{\rm g}$ for $\lambda_{\rm Edd}=0.05$, through
$36.62\,r_{\rm g}$ for  $\lambda_{\rm Edd}=0.1$ to $87.37\,r_{\rm g}$ for $\lambda_{\rm Edd}=0.3$. The heating ($\Gamma$) and cooling ($\Lambda$) rates corresponding to each transitional region are shown in the bottom panels of Fig.~\ref{fig:edding_ratio}. In all the cases, line cooling starts to dominate over Compton cooling, causing the warm corona temperature to drop rapidly.}

\begin{table}[]
\begin{center}
\caption{{   The properties of warm corona for three Eddington ratios and for Solar iron content. The edge of the warm corona and the width of transitional regions in units of $r_{\rm g}$ are given in columns 2 and 3. The dissipative heating at $r_{\rm edge}$ in $ 10^{-23}$ erg~cm$^{3}$s$^{-1}$, total SXE strength, and equivalent width (EW) of the iron line are listed in columns 4, 5, and 6, respectively.}}
\label{tab:wc_edd}
\begin{tabular}{c c c c c c }
\hline\hline
\noalign{\vskip 1.5mm}
 $\lambda_{\rm Edd}$ & $r_{\rm edge}\, [r_{\rm g}]$ & $\approx \Delta r$ &  
 $Q(r_{\rm edge}$) & $S^{\rm tot}_{\rm SXE}$ & Fe EW [eV]\\
 \noalign{\vskip 1.5mm}
\hline
   0.05 & 18.20 & 0.4 & $1.68 $ & 0.05 & 2732.7\\
   0.1 & 36.62 & 1.7 & $1.58 $ & 0.51 & 742.2 \\
   0.3 & 87.37 & 2 & $1.62 $ & 0.68 & 126.3 \\
\hline
\end{tabular}
\end{center}
\end{table}

\begin{figure}
    \centering
    \includegraphics[width=0.95\linewidth]{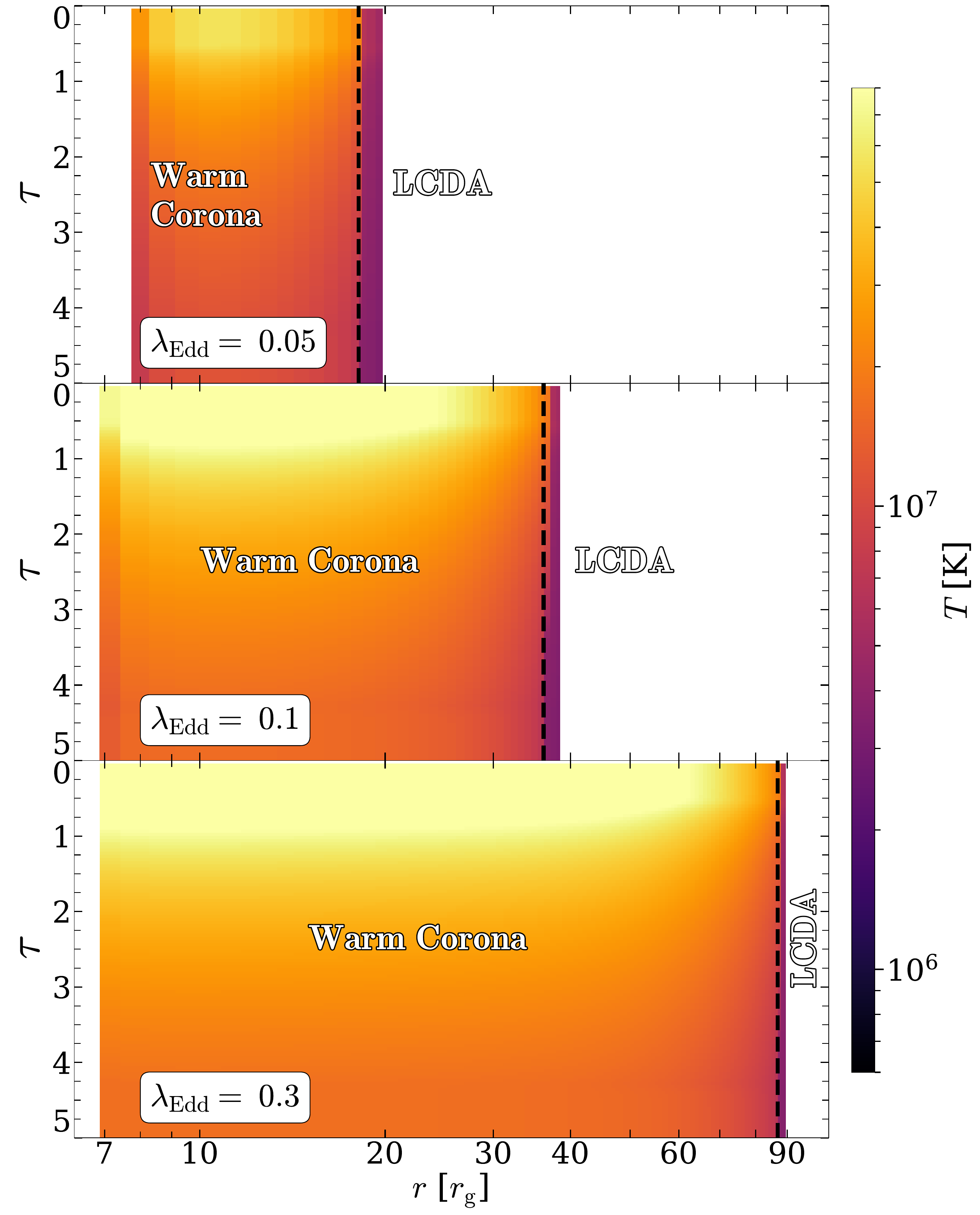}
   \caption{   The 2D temperature maps as a function of radius and optical depth, for the Eddington ratios given in the boxes. In each panel, we mark the warm corona and the line-cooled disk atmosphere (LCDA) separated by a vertical line at the location of the warm corona edge $r_{\rm edge}$.}
    \label{fig:edge_corona}
\end{figure}

{   For a better visualization of the warm corona size and edge, we present 2D temperature maps for all three values of $\lambda_{\rm Edd}$ in Fig.~\ref{fig:edge_corona}. The relative extent of the Comptonized atmosphere is clearly visible. It is worth noting that physical processes change gradually with radius, and the edge of a warm corona is not sharp. The width of the transitional region depends on thermodynamical and ionization properties. For $\lambda_{\rm Edd}=0.05$ it is small $\Delta r \approx 0.4\,r_{\rm g}$, whereas for $\lambda_{\rm Edd}=0.1$ it becomes wider $\Delta r \approx 1.7 \,r_{\rm g}$, reaching value   $\Delta r \approx 2\,r_{\rm g}$ in case of $\lambda_{\rm Edd}=0.3$. Beyond this region, only a line-cooled disk atmosphere (LCDA) remains. The radial extent of the warm corona and the transitional zone for three accretion rates are summarized in two columns of Tab.~\ref{tab:wc_edd}.}

\subsection{Effect of the Iron Abundance on the Warm Corona Edge} \label{sec:abundance}

\begin{figure*}
    \centering
    \includegraphics[width=1.0\linewidth]{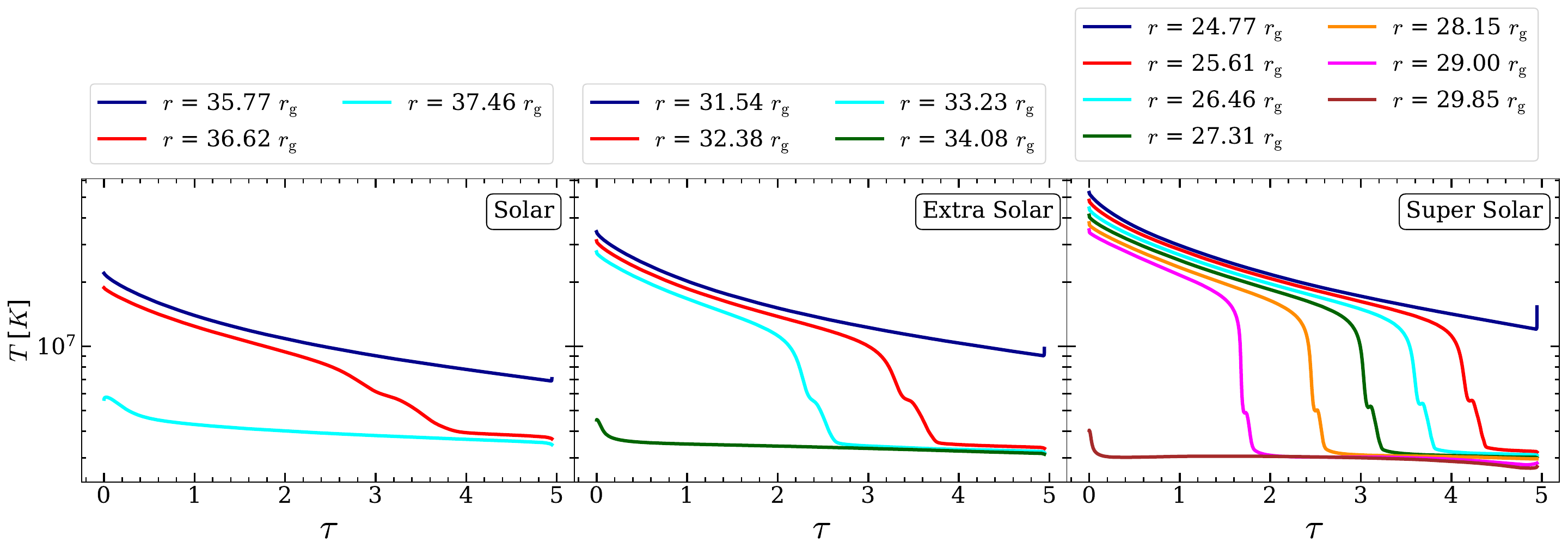}
    \caption{The temperature structures with optical depth for  various iron abundances, from left to right: 
    Solar, Extra Solar, and Super Solar, as listed in Tab.~\ref{tab:Fe_abund}, {   and for $\lambda_{\rm Edd}=0.1$.
    The different colors represent radial distances from the SMBH, depicted above each panel.}}
    \label{fig:abund_temp}
\end{figure*}

\begin{figure*}
    \centering
    \includegraphics[width=1.0\linewidth]{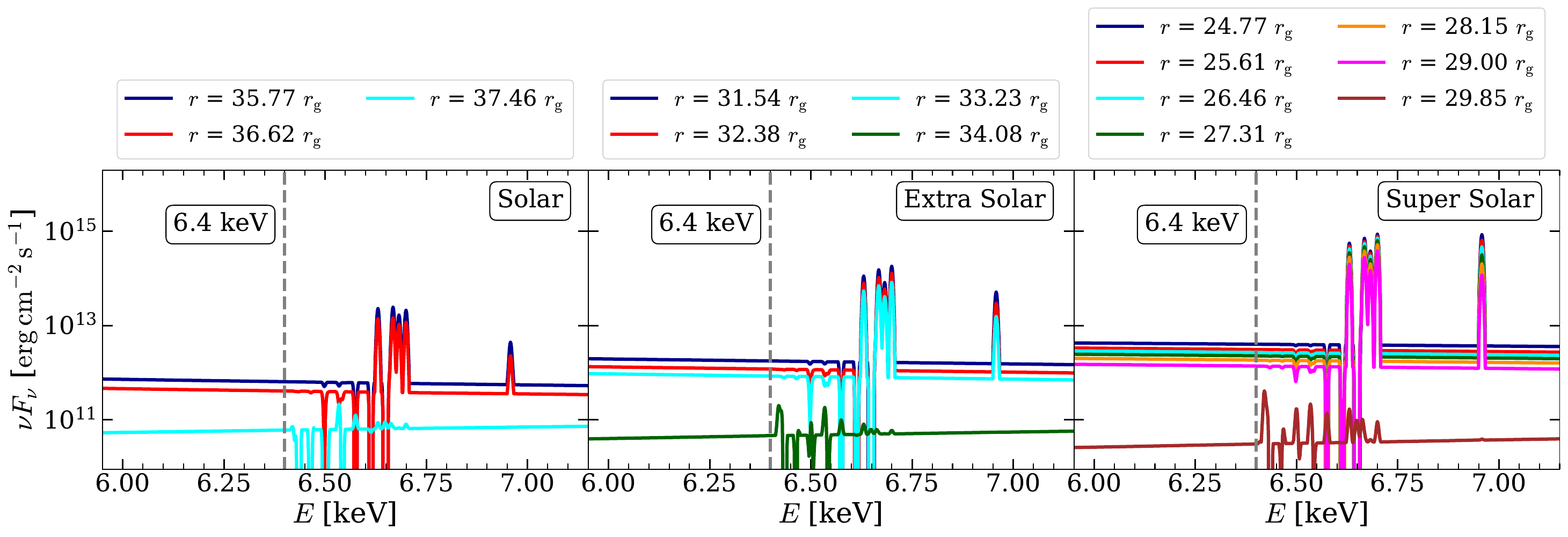}
    \caption{Angle-averaged reflected {   spectra produced by \texttt{TITAN} for
    various iron abundances, from left to right: Solar, Extra Solar, and Super Solar, as listed in Tab.~\ref{tab:Fe_abund}, and for 
    $\lambda_{\rm Edd}=0.1$. 
   The different colors represent the radial distances from the SMBH, depicted at each panel.}}
    \label{fig:abund_spec}
\end{figure*}

  {
 The line cooling process is closely tied to the chemical composition of the gas, and specifically to heavy elements, since hydrogen and helium are fully ionized in the illuminated accretion disk atmosphere \citep[][and references therein]{2008rozanska}. We also expect the lighter metals, such as C, N, and O, to be fully ionized in the warm corona, and we clearly demonstrate it in Appendix~\ref{app:elements}. The remaining metals, namely Ne, Mg, Si, and S, are almost fully ionized, and even where they are not, the thickness of their ionization fronts is much smaller than in the case of Fe, as illustrated in Fig.~\ref{fig:app_cno}.
 Because iron is the element that radiates the most energy through bound-free and bound-bound transitions, we vary its abundance to test the dependence of the warm corona edge on the metallicity.}

 \begin{table}[]
 \begin{center}
\caption{  {The iron content used for our simulations given as: 
ratio $A_{\rm Fe}$ relative to Solar - column 2, the ratio of iron to hydrogen number densities - column 3, and logarithmic absolute abundance - column 4. Columns 5 and 6 display the edge of the warm corona and the width of transitional regions for $\lambda_{\rm Edd}=0.1$.}
}
\label{tab:Fe_abund}
%\centering
\begin{tabular}{c c c c c c}
\hline\hline
\noalign{\vskip 1.5mm}
 Case & $A_{\rm Fe}$ & $n_{\rm Fe}/n_{\rm H}$ &  $\log(\varepsilon_{\rm Fe})$ & $r_{\rm edge}\, [r_{\rm g}]$ & $\approx \Delta r$ \\
\noalign{\vskip 1.5mm}
\hline
   Solar & 1 & $3.24 \times 10^{-5}$ & 7.51 & 36.62 & 1.7 \\
   Extra & 3.2 & $1.03 \times 10^{-4}$ & 8.01 & 32.38& 2.5\\
   Super & 10 & $3.24 \times 10^{-4}$ & 8.51 & 25.61 & 5.1 \\
\hline
\end{tabular}
\end{center}
\end{table}
 
 {   We consider three different iron abundances as 
  listed in Tab.~\ref{tab:Fe_abund}, keeping the same Eddington ratio $\lambda_{\rm Edd} = 0.1$. Fig.~\ref{fig:abund_temp} show the changes in the temperature structure with increasing metallicity. The rise of the temperature of the outermost atmospheric layers is clearly noticeable, naturally being explained} by the higher photoionization heating rate in iron-rich gas, driven by the growth of iron bound-free opacity \citep[][]{2013Garcia}. 

  {Furthermore, the overall shape of the temperature drop changes, and it is significantly different in the radial and vertical directions. 
  Vertical temperature drop becomes sharper with increasing iron content.} This follows directly from the line cooling mechanism: a higher iron abundance means more ions carrying L-shell electrons at a given depth, so once the gas recombines past Fe\,\textsc{xvii} the cooling rate rises much faster, and the thermal structure responds abruptly. {   Radially, the warm corona edge $r_{\rm edge}$ moves slightly inward, from $36.62\,r_{\rm g}$ for Solar iron content, through  $32.38\,r_{\rm g}$ for Extra Solar, and to $25.61 \,r_{\rm g}$ for Super Solar case. A factor of ten in iron abundance therefore shifts the edge by about $30\%$.
  The transitional region also broadens in comparison to the Solar case, as displayed in Tab.~\ref{tab:Fe_abund}. It becomes $\Delta r \approx 2.5 \,r_{\rm g}$ for Extra Solar metallicity, reaching $\Delta r \approx 5.1 \,r_{\rm g}$ in the case of Super Solar iron content. Hence, the dominant effect of the iron abundance in our models is therefore on the sharpness of the vertical ionization front, while the radial transitional region from the warm corona to a line-cooled disk atmosphere remains broader and therefore smoother.}

{   As a result, iron line features, occurring between 6.4--7~keV in the reflected spectrum, become more prominent for higher iron abundance. We illustrate it in Fig.~\ref{fig:abund_spec}, where angle-averaged local spectra produced by \texttt{TITAN} are plotted for the same radii as in Fig.~\ref{fig:abund_temp}. This is expected, since a higher metallicity leads to more ionic transitions and therefore stronger emission lines.
It is worth to notice that after the radial transition to the  {line-cooled disk atmosphere}, the neutral iron line near 6.4~keV also becomes stronger with increasing iron content.}

\subsection{Observational signatures of SXE}
\label{subsec:obser}

To illustrate the effect of a warm corona size {   on the total emitted spectrum}, we built a complete emission model using \texttt{TiNoGyo}, as described in Sec.~\ref{subsec:titan}, for three values of $\lambda_{\rm Edd}~=~0.05$, $0.1$, and $0.3$, and Solar iron content. We point out here that the total spectrum {   discussed by us in this paper} does not include a power-law emission from the hot corona, usually fitted to the observed data. We study only two components here: the reflection from the optically thick, dissipative {   disk atmosphere} above a disk, and the SXE originating in the warm corona itself.

\begin{figure}
    \centering
    \includegraphics[width=1.0\linewidth]{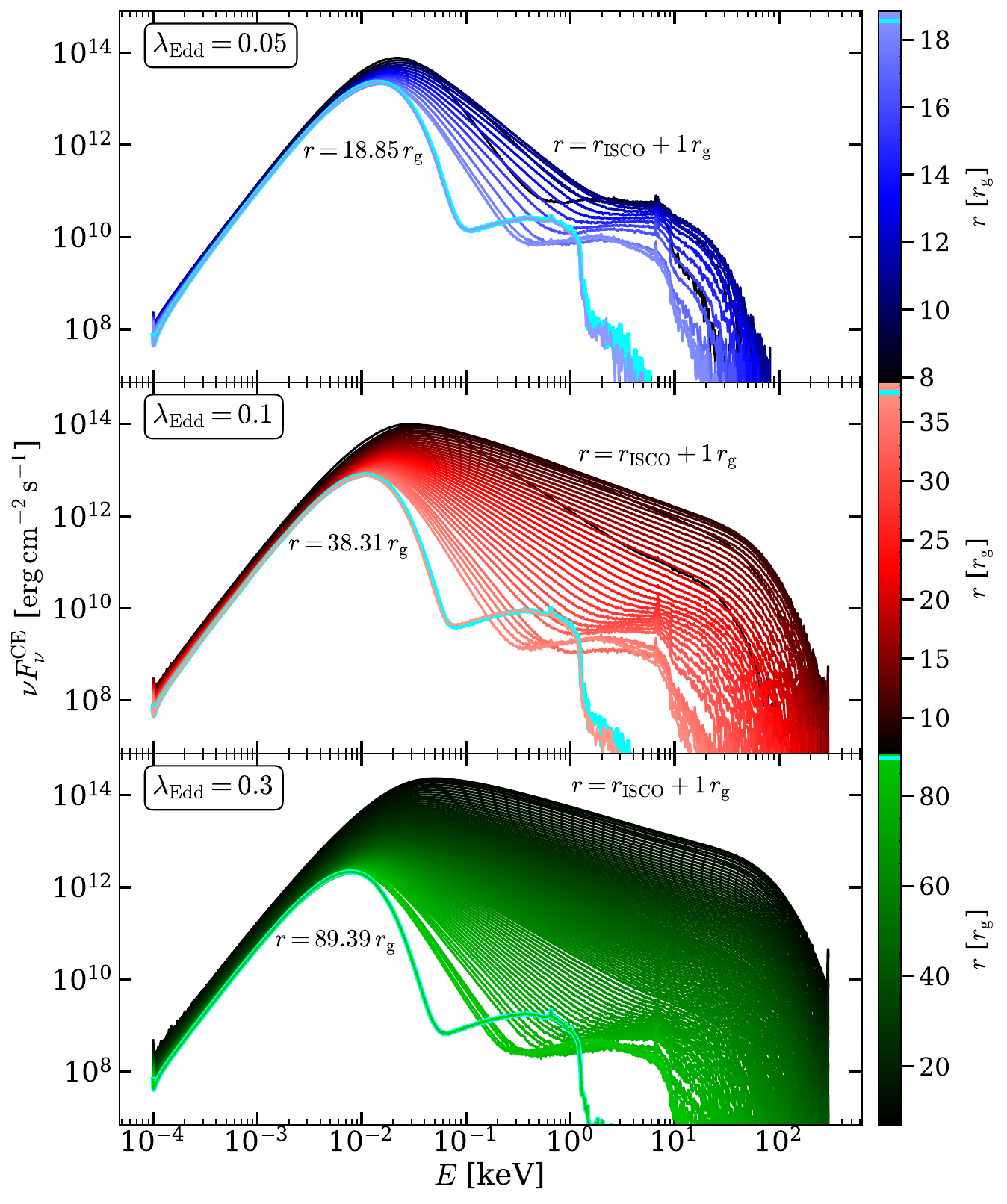}
    \caption{The outgoing spectra of Compton-upscattered photons computed with \texttt{NOAR} for the Eddington ratios shown in the boxes. {   Each spectrum is emitted locally at a given radial distance} from the SMBH, shown in the color bar.  The cyan line marks the spectrum from the LCDA.}
    \label{fig:edding_spec_trans}
\end{figure}

 {The Comptonized spectra produced by \texttt{NOAR} locally at radii spanning from $r_{\rm ISCO}+1\,r_{\rm g}$ up to the radius where LCDA starts are plotted in  Fig.~\ref{fig:edding_spec_trans} for three accretion rates. 
 Spectra emitted from radii lower than $r_{\rm edge}$, are clearly Comptonized and extend even into the hard X-ray band for the highest $\lambda_{\rm Edd}$. While moving across the transitional region, spectra become softer and less Comptonized. At the radius where the atmosphere becomes predominantly line-cooled, the emitted spectrum given by the cyan line possesses a clear cut-off at 1~keV.
 It is already evident that the largest contribution to the Comptonized emission comes from the warm corona with the largest radial extent, i.e., the one obtained for
  $\lambda_{\rm Edd}=0.3$.}

\begin{figure}
    \centering
    \includegraphics[width=0.95\linewidth]{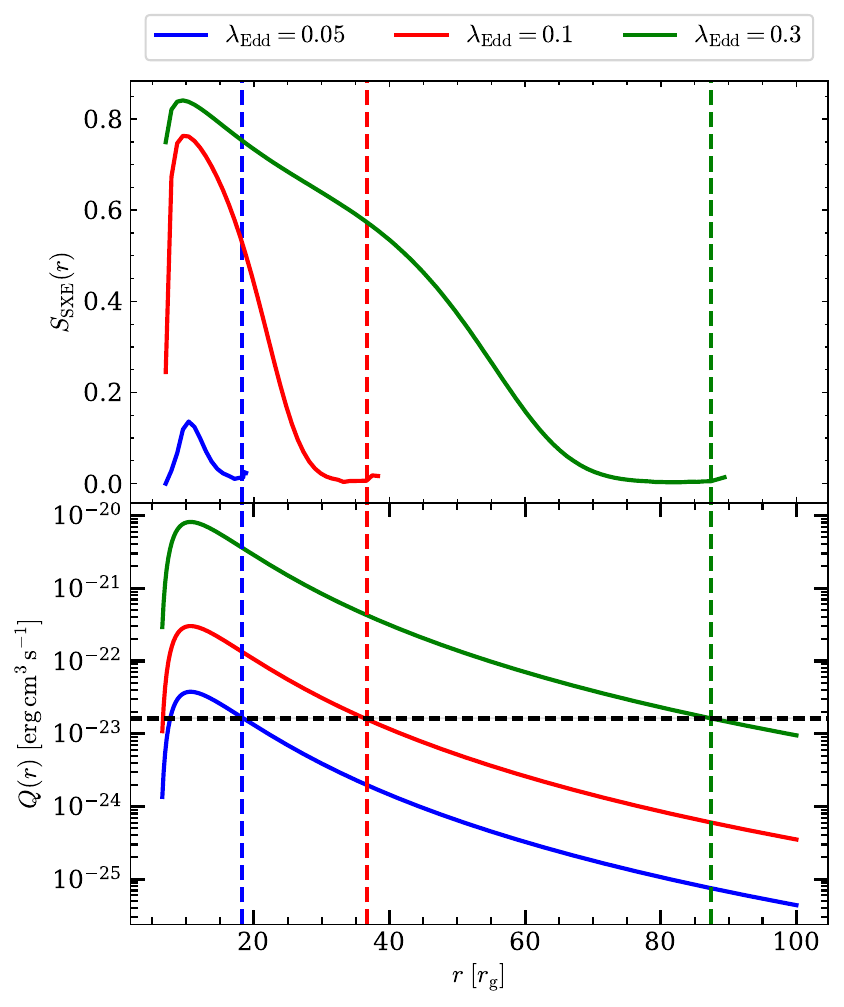}
    \caption{ {{\it Upper panel:} The SXE strength as a function of radius for $\lambda_{\rm Edd}=0.05$ (blue), $0.1$ (red), and $0.3$ (green). The dashed vertical lines mark $r_{\rm edge}$ in the color of the corresponding $\lambda_{\rm Edd}$. {\it Bottom panel:} The dissipative heating of an accretion disk atmosphere across the size of a warm corona. Color lines have the same meaning as in the upper panel.  Black horizontal line crosses the three intersection points, showing that the heating rate at the edge of the warm corona is the same in all three cases. }}
    \label{fig:SXE_strength}
\end{figure}

\begin{figure}
    \centering
    \includegraphics[width=1.0 \linewidth]{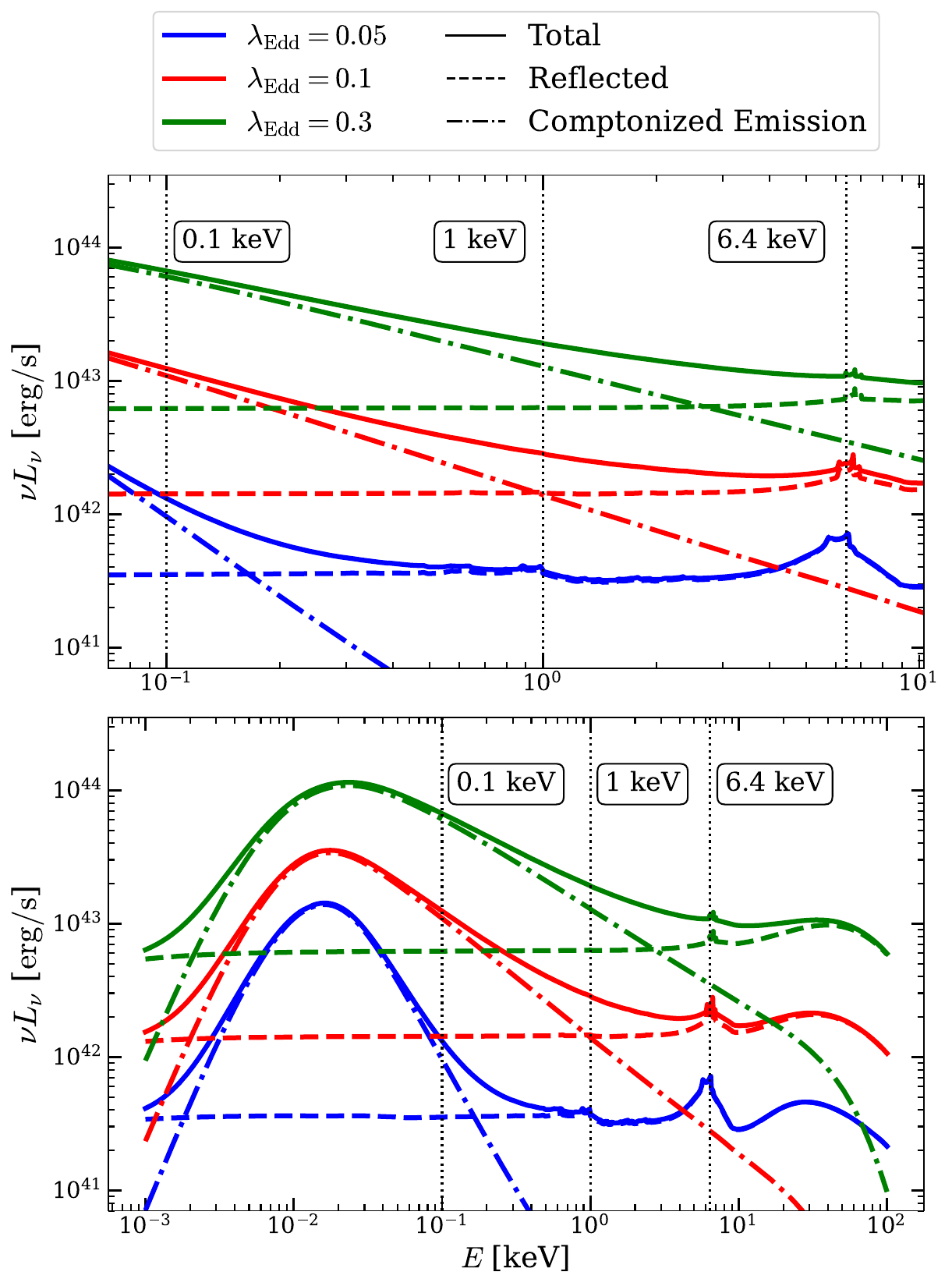}
    \caption{The total observable spectra generated from \texttt{TiNoGyo} for the face-on observer and
     different Eddington ratios: blue for 0.05, red for 0.1, and green for 0.3.  The solid line is the total spectrum, which is the sum of the reflected component (as in the typical lamp-post model, dashed line) and the Comptonized warm corona component (dashed-dotted line). The top panel shows the 0.07-10~keV energy range, and the bottom panel shows the broadband spectral energy distribution.}
    \label{fig:edding_spec}
\end{figure}

\begin{figure}
    \centering
    \includegraphics[width=1.0 \linewidth]{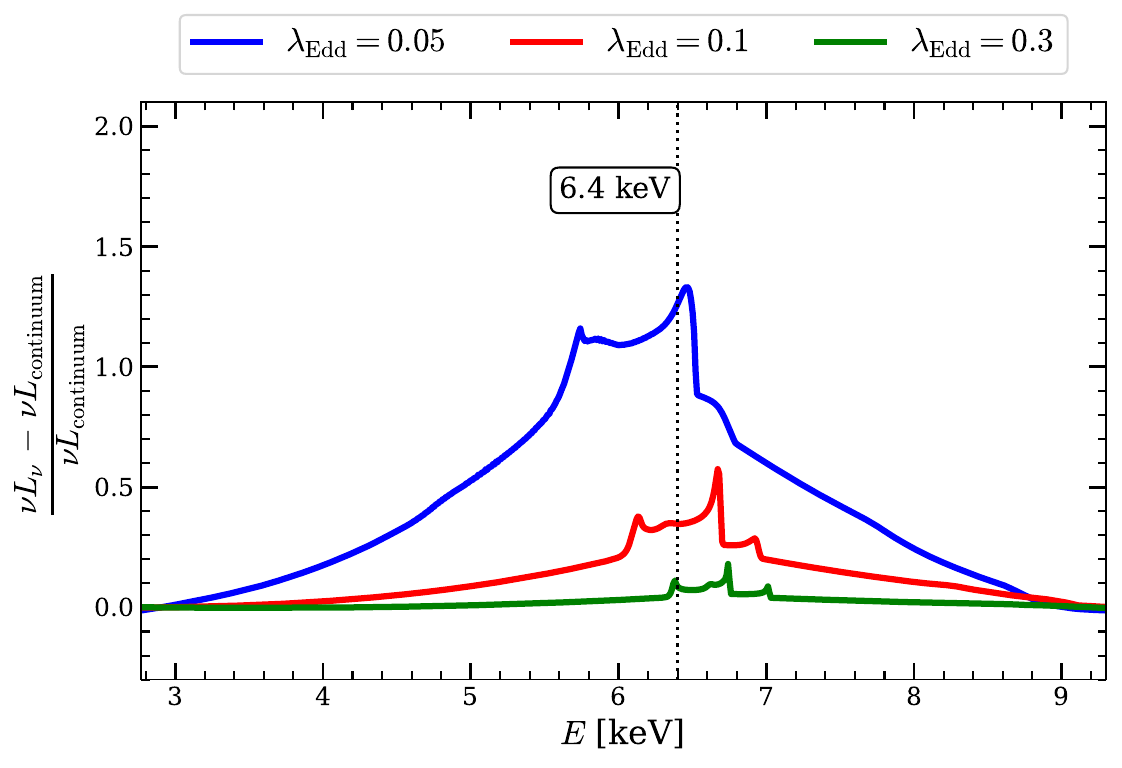}
    \caption{The continuum-reduced spectra near the iron line profile, shown between 3-9~keV. Different colors represent variations of the Eddington ratio, same as in Fig.~\ref{fig:edding_spec}. }
    \label{fig:edding_spec_irl}
\end{figure}

 {To verify  that the warm corona emission indeed terminates at $r_{\rm edge}$, we compute the SXE strength according to Eq.~\ref{eq:SXE_strength}.
The resulting radial profiles of $S_{\rm SXE}(r)$ are shown in the upper panel of Fig.~\ref{fig:SXE_strength}, where vertical lines mark the position of $r_{\rm edge}$ for different $\lambda_{\rm Edd}$. In all cases, $S_{\rm SXE}(r)$ rises to a maximum near $r \approx 10\,r_{\rm g}$ and then decreases outward, becoming negligible at $r_{\rm edge}$. The above behavior follows the radial stratification of the dissipative heating $Q(r)$ of the atmosphere, which we plot in the bottom panel of Fig.~\ref{fig:SXE_strength}. 
Crucially, the whole region producing the SXE lies inside $r_{\rm edge}$, which confirms that our criterion yields a conservative estimate of the warm corona size.
Most importantly, the value of the heating rate at the edge of the warm corona defined by us is identical for the three accretion rates, as shown by the horizontal black line in Fig.~\ref{fig:SXE_strength}. 
In summary, our definition of the corona edge, where Compton cooling transitions to line cooling, is justified and takes place for a specific heating value depicted in the fourth column of Tab.~\ref{tab:wc_edd}.}

The final observable spectra from the full \texttt{TiNoGyo} procedure, after integration over the emission surface and including all general relativistic effects, are shown in Fig.~\ref{fig:edding_spec}. The total spectrum is a sum of reflected emission given by the dashed line and Comptonized emission from a warm corona plotted by the dashed-dotted line. The deviation of the total spectrum from the reflected component in the soft X-ray band is clearly due to the Comptonized emission {   originating from a warm corona.
The extent of the warm Comptonization towards harder X-rays
increases with Eddington ratio, as a consequence of the larger warm corona size.}
 {The observed trend of increasing SXE strength with increasing $\lambda_{\rm Edd}$ arises from both the radial extent of the warm corona and the higher $S_{\rm SXE}$  value at each radius, as demonstrated in Fig.~\ref{fig:SXE_strength}.}

The warm corona emission is featureless. The only feature visible in the total spectrum is the iron line complex.
{   To investigate how these lines depend on the accretion rate after the ray-tracing procedure,}
 we present continuum-reduced emission spectra around the 6.4~keV region in Fig.~\ref{fig:edding_spec_irl}. The most prominent, relativistically distorted iron line complex is observed at the lowest considered $\lambda_{\rm Edd}$.  {To quantify this result, we compute the equivalent width (EW) of the line complex over 3--9~keV, taking the underlying continuum as a total Comptonized plus reflected spectrum. The EWs values are listed in the last column of Tab.~\ref{tab:wc_edd}. For small accretion rate, the warm corona is compact and bound-bound transitions on iron ions become efficient close to SMBH, producing a prominent, broad iron line complex.} However, for $\lambda_{\rm Edd} = 0.3$, the radial extent of the warm corona increases {   by a factor of about 4.5}, giving rise to two distinct effects: i) surface temperature at smaller radii closer to SMBH is larger, so the iron is completely ionized and does not produce lines, ii) the line originates from a cooler atmosphere at outer radii, but the profile is only shaped by Keplerian motion and not affected by the relativistic effects due to the large distance from the SMBH.

{   }
 
 {The aim of our work was to explain the origin of the soft X-ray excess and how its emission relates to the reflected spectrum.
Thus, spectra presented in Figs.~\ref{fig:edding_spec} and~\ref{fig:edding_spec_irl} include emission from radii up to $r_{\rm edge}$, where our grid ends.
But the reflection from the disk atmosphere at larger radii, examples of which can be seen in the local spectra presented in Fig.~\ref{fig:abund_spec}, may also contribute to the overall spectrum. 
 Since relativistic broadening is weak at such radii, the distant reflection would likely add a narrow, neutral component near 6.4~keV, which we expect to matter most at low $\lambda_{\rm Edd}$, where the warm corona spans the smallest range of radii. Due to the complexity of our model and the long computational time, we address this issue in future work.}

%%%%%%%%%%%%%%%%%%%%%%%%%%%%%%%%%%%%%%%%%%%%%%%%%%%%%%%%%%%%%%
\section{Discussion and Conclusion}
\label{sec:discon}

{   We developed a model of a warm corona as an optically thick, dissipative atmosphere located on top of a standard accretion disk, in a two-slab geometry. Radially stratified input was constructed \citep{2022Xiang,2026Biswas}  in order to compute the vertical radiative transfer with the
\texttt{TiNoGyo} pipeline. 
We explored in detail the case of non-rotating SMBH for three different accretion rates $\lambda_{\rm Edd} = 0.05, 0.1, 0.3$, and three iron abundances.

We found that the outer edge of the warm corona forms where Compton cooling is replaced by line cooling, which is marked by a sudden drop in gas temperature. Beyond this edge, only the line-cooled disk atmosphere (LCDA) remains, with no warm corona present.} Several earlier studies have pointed to line cooling as an important process  \citep[see for instance][]{2001Ballantyne, 2013Garcia}, but we built a complete model to demonstrate that the line cooling sets the outer edge of the warm corona.
{   Moreover, we showed that the dissipative heating rate 
plays a key role: at the edge of the warm corona, it reaches the same value for all accretion rates, i.e., $Q(r_{\rm edge}) \approx 1.6 \times \times10^{-23}$ erg\,cm$^{3}$\,s$^{-1}$ for Solar chemical composition. This is related to the shape of the function  $Q(r) \propto\lambda_{\rm Edd}^{3}\,r^{-9/2}J(r)^{3}$.}

{   With such a defined transitional region from Comptonized warm corona to LCDA, we demonstrated that the warm corona size increases with increasing accretion rate, consistent with recent observations of the AGN sample \citep{2024Palit}}, and with numerical calculations of magnetically supported warm corona and accretion disk structure by \citet{2023Gronkiewicz}. 

{   As a consequence, the strength of SXE computed from synthetic spectra produced by the combination of photoionization, Comptonization and ray-tracing codes \texttt{TiNoGyo} grows with $\lambda_{\rm Edd}$. 
At higher Eddington ratios, the warm corona edge is located at a larger radius, leading to a larger emitting area, and consequently domination of the Comptonized component over the reflected spectrum around 1~keV. This trend is also supported by recent observations: the CLAGN Mkn~590 showed an increase in SXE emission following a rise in the Eddington ratio \citep{2025palit}, which can be explained by the increase in the warm corona size.}

 {The line cooling directly depends on the gas chemical composition, but the strongest and most wide ionization fronts come from iron (Appendix~\ref{app:elements}). 
Therefore, by testing metallicity, we considered only the iron content as a representative element that has the strongest impact on line cooling \citep{2000madej}.
Higher iron abundance acts mainly on the vertical structure, steepening the temperature drop across the ionization front. Its effect on the radial extent of the warm corona is comparatively weak: a factor of ten in iron abundance moves $r_{\rm edge}$ inward by about $30\%$, and the radial transition to a line-cooled disk atmosphere becomes smoother rather than sharper.}

{   The Comptonized continuum from the warm corona itself is featureless, but we can still extract iron-line profiles originating from the reflected spectrum. The width of the ray-traced iron line depends on $\lambda_{\rm Edd}$: at high $\lambda_{\rm Edd}$, iron near the SMBH is fully ionized, so the Fe emission line complex forms at large radii only weakly affected by relativistic effects. At low $\lambda_{\rm Edd}$, the smaller warm corona results in the Fe line forming closer to the black hole, giving a larger redshift and a profile that extends well above the continuum. The exact EW values will change once the illumination by the power-law continuum from a hot corona is included, but their relative values will remain unchanged.}

{   The major limitation of our model is that we do not continue \texttt{TiNoGyo} computations}
 beyond the transition from the warm corona to the  {LCDA even if the temperature remains higher than the standard accretion disk given by \citet{1973Shakura}}. We expect that further out  {radially, the ionization fronts from other elements, from oxygen to sulfur, will act as efficient coolants once the temperature drops enough for them to recombine  {(see Appendix~\ref{app:elements})}}, potentially bringing the  {disk atmosphere} closer to temperatures typical of a standard disk. The line cooling from atomic processes is usually neglected in GRRMHD (General Relativistic Radiation Magnetohydrodynamics) simulations of accretion around SMBHs, but turns out to be essential for modeling the warm corona in AGN \citep[see ][]{2019Jiang}. This sets AGN apart from X-ray binaries, where the disk is hot enough for Compton and Bremsstrahlung cooling to dominate. %This picture is further supported by the findings of \citet{2026Zdziarski}, who proposed that an additional warm corona component can reconcile the spins of binary black hole mergers with those measured in X-ray binaries.

 {In this work, the optical depth of an atmosphere is fixed over radii, but it enters the internal heating directly $Q \propto 1/\tau$ (Eq.~\ref{eq:inputs}). Thus, for a given dissipated flux, a larger $\tau$ spreads the same energy over more material, lowering the value of local $Q$. As shown in Sec.~\ref{subsec:obser}, the warm corona terminates where the dissipative heating falls below a critical threshold.
For a given $\lambda_{\rm Edd}$, a higher optical depth would move $r_{\rm edge}$ inward, and therefore partly offset the outward shift we find at constant $\tau$. Whether the net trend survives depends on how steeply $\tau$ grows with $\lambda_{\rm Edd}$, which our present simulations cannot constrain. We defer the joint variation of $\tau$ and $\lambda_{\rm Edd}$, together with the dependence on SMBH spin, to a wider parameter grid in future work.}

{   The radial extent of the warm corona also scales with the iron abundance of the accretion flow.} We assume Solar metallicity, consistent with the early enrichment of galaxy centers inferred from observations \citep{2022Wang}.
Another point worth noting is the energy budget used in all our simulations: 20\% of the total flux comes from the X-ray lamp illumination, 40\% from the dissipative warm corona, and 40\% from the inner accretion disk. Changing this distribution may affect where the transition occurs, which we also leave to future work. We also plan to test our models against data from future missions such as NewAthena.

%%%%%%%%%%%%%%%%%%%%%%%%%%%%%%%%%%%%%%%%%%%%%%%%%%%%%%%%%%%%%%
\begin{acknowledgments}
PPB and RW have been fully supported, and AR has been partially supported by the Polish National Science Center grant No. 2021/41/B/ST9/04110. All computations have been performed using the computer cluster at NCAC PAS. DL
acknowledges the Czech Science Foundation (GA\v{C}R) grant no. 25-16928O. AAZ acknowledges support from the Polish National Science Center grants 2023/48/Q/ST9/00138 and 2025/59/B/ST9/01180. We thank the referee for constructive comments that improved this paper.
\end{acknowledgments}

\bibliography{edge_wc_parik}{}
\bibliographystyle{aasjournalv7}

\appendix
\section{The impact of metals on the line cooling rate}
\label{app:elements}

\begin{figure*}
    \centering
    \includegraphics[width=0.48\linewidth]{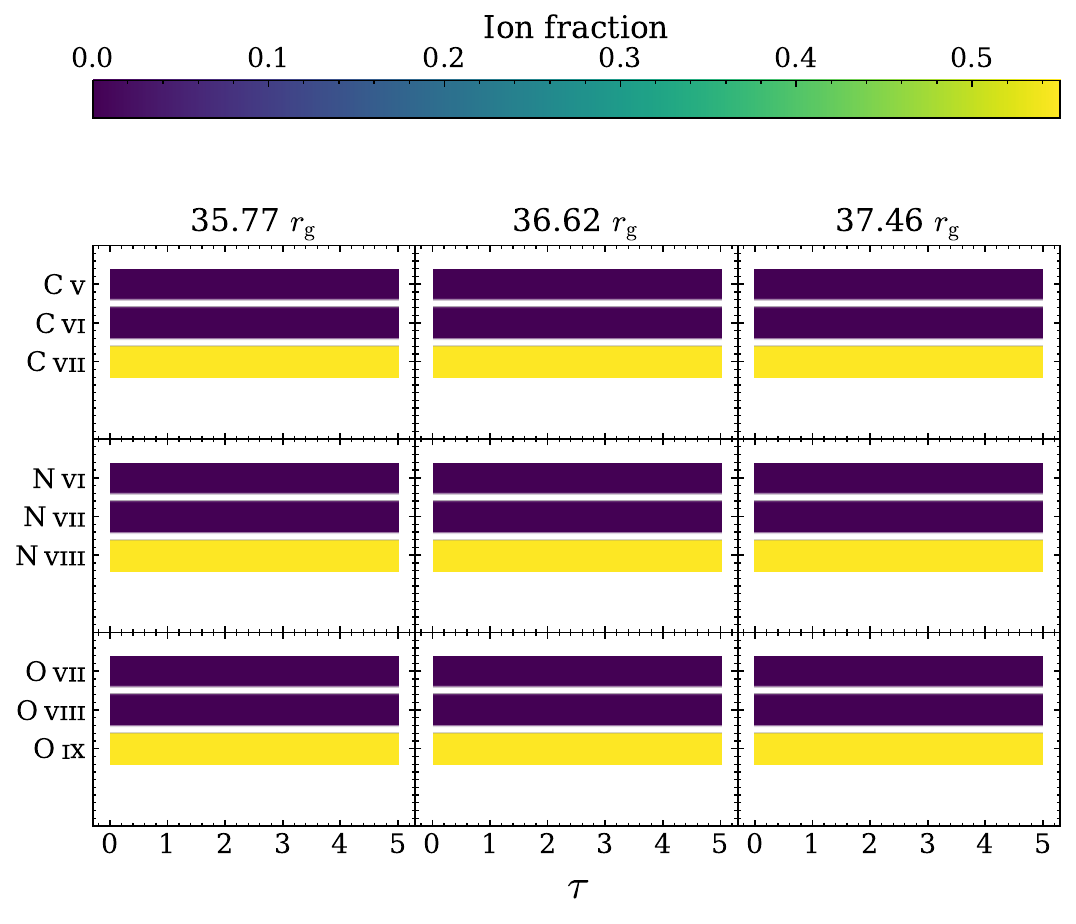}
    \includegraphics[width=0.48\linewidth]{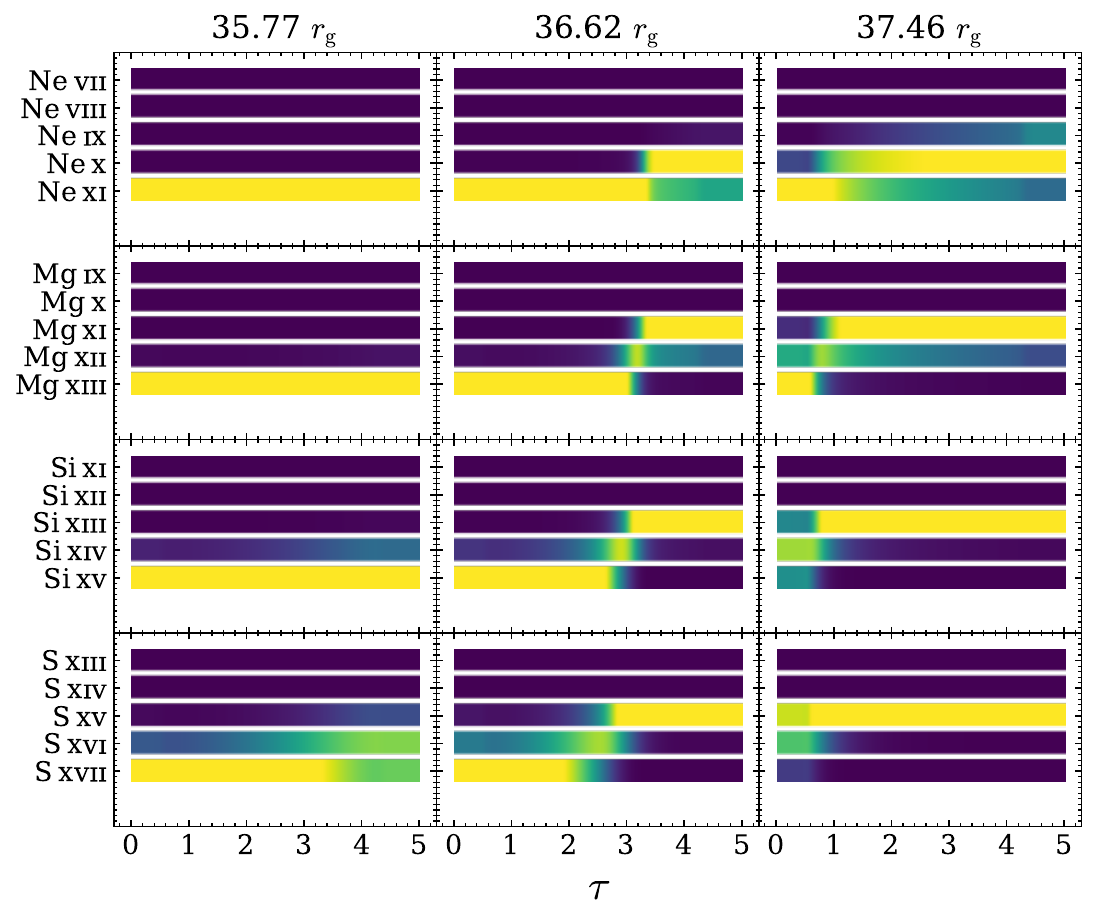}
    \caption{{\it Left panel}: Ion fractions of C, N, and O (from top to bottom panel respectively) as a function of optical depth given by the color scale in the side bar. The results of \texttt{TITAN} at the three radii marked above the panels, 
    bracketing the transition region for $\lambda_{\rm Edd}=0.1$ and Solar abundance. The ionization degree is shown on the left side of each panel. For light metals, the entire population resides in the fully stripped ion, with no change across the transition. {\it Right panel}: The same but for Ne, Mg, Si and S. Narrow ionization fronts are visible, but their width is significantly smaller than in the case of Fe shown in Fig.~\ref{fig:hc_spin_0}.}
    \label{fig:app_cno}
\end{figure*}

Chemical composition has a major impact on the line cooling rate, and the \texttt{TiNGyo} procedure allows to trace it by changing metallicity in the photoionization code \texttt{TITAN}. The abundance of 10 elements can be varied,
but this introduces ten additional free parameters, whose individual effects are difficult to disentangle. The \texttt{TITAN} output gives the net line cooling rate as a total over all transitions and does not decompose it by element.
Therefore, to show how the size of the warm corona depends on the chemical composition, we use iron alone as a proxy for metallicity, since iron contributes most to the energy exchange via bound-free and bound-bound transitions.
Nevertheless, as in any photoionization code, the ionization fronts of different elements can be identified in \texttt{TITAN} from their ionization structure. It is obvious that hydrogen and helium are fully ionized, thus, we omit them from the discussion. 

The left plot in Fig.~\ref{fig:app_cno} shows the ion fractions of carbon, nitrogen, and oxygen at the three radii, which enclose the transitional region for $\lambda_{\rm Edd} = 0.1$ and Solar abundances, as displayed in Fig.~\ref{fig:hc_spin_0}. In all three cases, the population resides entirely in the fully stripped ions, C\,\textsc{vii}, N\,\textsc{viii}, and O\,\textsc{ix}, over the whole optical depth. Thus those metals retain no bound electrons anywhere in the region of interest and cannot contribute to line cooling. Their ionization structure is unchanged across the transition, so they cannot produce a sharp temperature drop occurring at $r_{\rm edge}$.

%and Fig.~\ref{fig:app_sisu}
Neon, magnesium, silicon, and sulfur do recombine across the transition, as shown in the right panel of  Fig.~\ref{fig:app_cno}. At $r=35.77\,r_{\rm g}$ they are fully stripped. At $r=36.62\,r_{\rm g}$ they begin to recombine below $\tau\approx3$, and by $r=37.46\,r_{\rm g}$ the dominant ionization states are Ne\,\textsc{x}, Mg\,\textsc{xi}, Si\,\textsc{xiii}, and S\,\textsc{xv}. These are hydrogen-like and helium-like configurations, so their only bound electrons are in the K-shell, for which line cooling is known to be inefficient (see Sec.~\ref{sec:res}). 

Iron is the only element in our composition that acquires a filled L-shell at the transition, shifting from the helium-like Fe\,\textsc{xxv} to the neon-like Fe\,\textsc{xvii} with ten bound electrons (Fig.~\ref{fig:hc_spin_0}, bottom panels). It therefore supplies an efficient line cooling channel, which serves as the basis for our definition of $r_{\rm edge}$. 
This situation is specific to the temperatures reached in the warm corona and just beyond its edge. Further out along the line-cooled disk atmosphere, where the temperature continues to drop, the lighter elements will in turn recombine and their own ionization fronts are expected to appear.

\end{document}